\documentclass[11pt]{iopart}
\usepackage{graphicx}
\usepackage{subfigure}
\usepackage{morefloats}
\usepackage{color}
\usepackage{setspace}
\usepackage{floatrow}
\usepackage{amsmath,bm}
\begin{document}

\title{Universality of the ion flux to the JET outer wall}
\author{ N. R. Walkden$^{1}$, A. Wynn$^{1,2}$, F. Militello$^{1}$, B. Lipschultz$^{2}$, G. Matthews$^{1}$, C.Guillemaut$^{1}$, J. Harrison$^{1}$, D. Moulton$^{1}$ and JET Contributors${*}$
		\\ \small{EUROfusion Consortium, JET, Culham Science Centre, Abingdon, OX14 3DB, UK}
        \\ \small{$^{1}$ CCFE, Culham Science Centre, Abingdon, OX14 3DB, UK} 
        \\ \small{$^{2}$ York Plasma Institute, Department of Physics, University of York, Heslington, York, YO10 5DD, UK} 
		\\ \small{$^{*}$See the Appendix of F.Romanelli et al, Proceedings of the 25th IAEA Fusion Energy Conference 2014, Saint Petersburg, Russia}
        \\ Email: \texttt{nick.walkden@ccfe.ac.uk} }
\date{}

\begin{abstract}
Universality in the ion flux to the JET outer-wall is observed in outer-wall limiter mounted Langmuir probe (OLP) time-series across a large range of plasma current and line-averaged density during Ohmically heated horizontal target L-mode plasmas. The mean, $\mu$, and the standard deviation, $\sigma$, of the ion-saturation current measured by the OLP show systematic variation with plasma current and density. Both increase as either plasma current decreases and/or density increases. Upon renormalization, achieved by subtraction of $\mu$ and rescaling by $\sigma$, the probability distribution functions (PDFs) of each signal collapse approximately onto a single curve. The shape of the curve deviates from a $\Gamma$ distribution in the tail of the PDF and is better described by a log-normal distribution. The collapse occurs over 4 decades of the ordinate which, given the wide parameter space over which the data spans, is a strong indication of universality. The invariance in the shape of the PDF is shown to be the result of a balance between the duration time of the average burst wave-form, $\tau_{d}$ and the waiting time between bursts, $\tau_{w}$. This implies that the intermittency parameter, $\tau_{d}/\tau_{w}$, can be considered constant at the JET outer wall during horizontal target Ohmic L-mode operation. This result may be important both for model validation and prediction.
\end{abstract}

\section{Introduction}
Protection of first-wall and divertor components in future reactor scale tokamaks will be critical \cite{WenningerNF2015}. Excess fluxes of particles and heat to the first wall can cause errosion and sputtering leading to damage, Tritium retention and/or dust production \cite{RothJNM2009}. Particle loss to the first-wall affects the steady state particle balance of the plasma impacting global factors such as fuelling and detachment \cite{GuillemautNF2014} so understanding the cross-field transport processes responsible for this loss is essential. The bulk of particles that hit the first wall are carried in intermittent busty events called blobs or filaments that are ejected from the core plasma and propagate coherently through the scrape-off layer (SOL) \cite{D'IppolitoReview}. Single-point time-series taken on Langmuir probes \cite{GarciaNF2015,HnatNF2008,GravesPPCF2005,VanMilliganPoP2005} or with optical diagnostics \cite{GarciaJNM2013,GarciaPoP2013} show skewed and lepto-kurtic probability distribution functions (PDFs) as a result of these bursty events. Recent theoretical progress has been made in describing the features of these time-series by modelling them as the result of a shot noise process; a series of uncorrelated random bursts with a fixed waveform and exponentially distributed amplitudes and waiting times \cite{GarciaPRL2012}. This model not only captures the shape of the PDFs, which are found to be described by a Gamma distribution in agreement with early experimental observations \cite{GravesPPCF2005}, but predicts the commonly observed quadratic relation between the skewness (3rd order moment of the PDF) and kurtosis (4th order moment) \cite{SattinPPCF2009}. Large datasets taken with probe measurements on TCV and GPI measurements on Alcator C-mod have solidified the basis for this model\cite{GarciaNF2015,GarciaPoP2013}. Extensions to this stochastic model now permit predictions of radial SOL profile shapes based on this statistical theory \cite{Militello2016ArXiV1,Militello2016ArXiV2}. 
\\The behaviour of time-series in the SOL of tokamaks and other magnetic confinement devices is remarkably universal across plasma conditions \cite{GravesPPCF2005} and even across devices \cite{VanMilliganPoP2005}. In TCV the PDF of the ion saturation current time series measured 7mm into the SOL was shown to collapse under re-scaling across different densities, plasma currents and even confinement modes \cite{GravesPPCF2005}. Furthermore on Alcator C-mod GPI measurements in the far SOL \cite{GarciaJNM2013,GarciaPoP2013} and JET reciprocating probe measurements close to the separatrix \cite{HidalgoPPCF2002} a similar PDF collapse was observed. In this paper we further investigate the universality of ion-saturation current time-series however this time focussing on measurements taken at the plasma wall using an outer wall-mounted langmuir probe (OLP) on JET. We conduct our analysis across a range of line-averaged plasma densities from $1.69 - 3.93 \times 10^{19}m^{-3}$ and a range of plasma currents from $1.5$ to $3$MA (to our knowledge the widest scan in $I_{p}$ thus far considered) which correspond to a range of Greenwald fractions from $0.1$ to $0.7$. All plasmas are Ohmically heated in horizontal target configuration. The letter is organised as follows: Section \ref{Sec:Exp} describes the experimental setup used for this analysis and discusses the appropriateness of the data for statistical analysis. Section \ref{Sec:Res} describes the statistical analysis performed on the timeseries described in section \ref{Sec:Exp} before section \ref{Sec:Conc} discusses the results and concludes.

\section{Experimental Setup}
\label{Sec:Exp}
\begin{figure}[htbp]
\centering
\includegraphics[width=0.8\textwidth]{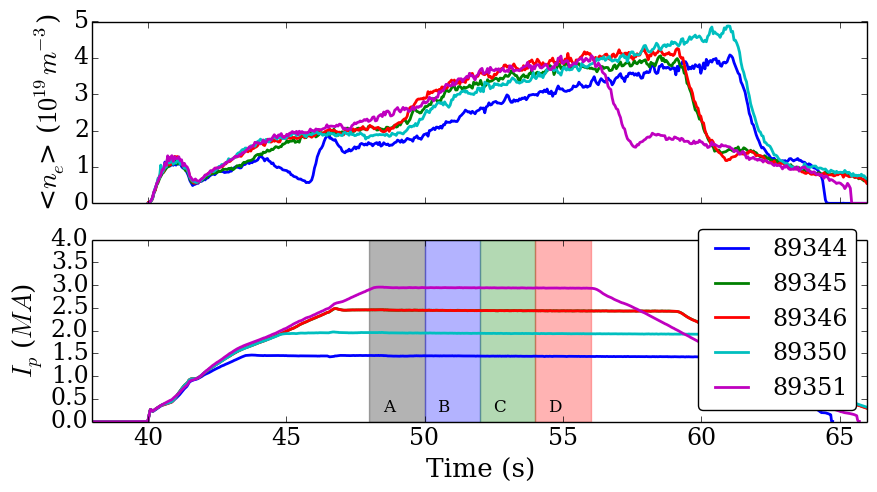}
\caption{Line averaged plasma density (upper) and plasma current (lower) for the five pulses analysed. Also highlighted are the four windows, labelled A, B, C and D respectively, that the analysis described herein is carried out on.}
\label{Fig:Params}
\end{figure}

\begin{figure}[htbp]
\centering
\includegraphics[width=0.3\textwidth]{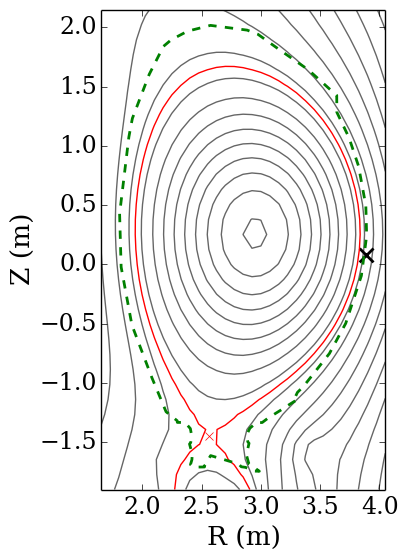}
\caption{Equilibrium magnetic flux surfaces for pulse 89344 at $52$s. Overlaid is the separatrix position (red solid line), the first-wall (dashed green line) and the position of the wall-mounted Langmuir probe used in this study (black cross).}
\label{Fig:Probe_pos}
\end{figure}

\begin{figure}[htbp]
\centering
\includegraphics[width=\textwidth]{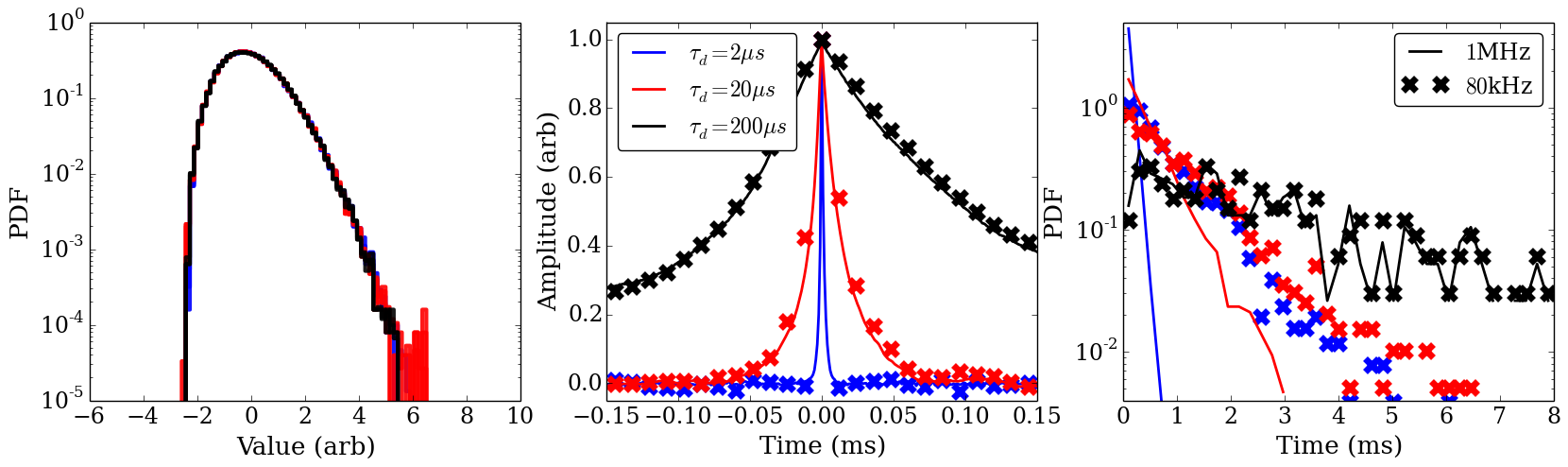}
\caption{Left: PDFs of the signals, with the mean subtracted and rescaled by the standard deviation. All values of $\tau_{d}$ at each sampling frequency are shown, however the overlap between curves makes them indistinguishable. Center: Conditionally averaged waveform for each value of $\tau_{d}$ for the 1MHz (solid lines) and 80kHz (crosses) signals. Right: Waiting time distribution for each value of $\tau_{d}$ for the original signal (solid lines) and the 80kHz sampled signal (crosses).}
\label{Fig:synth_pdfs}
\end{figure}

The data analysed in this letter was taken during the JET C35 experimental campaign during pulses $89344,89345,89346,89350$ and $89351$.  Figure \ref{Fig:Params} shows time traces of line averaged plasma density and plasma current during the time window analysed in this paper. Each pulse contained a ramp in line averaged density. The time window $48$s to $56$s (flat-top phase) is split into four sub-windows of $2$s providing a total of 20 individual time-series' to be analysed.  These windows are highlighted and lettered in figure \ref{Fig:Params}. We note that the repeated pulse at $I_{p} = 2MA$ (89345 and 89346) demonstrates the repeatability of this process.
The toroidal magnetic field and plasma current were varied in a manner designed to keep the connection length in the SOL approximately constant between pulses.
\\The measurement under focus in this letter is the ion-saturation current ($J_{sat}$) measured on an OLP situated just below the outboard midplane. Figure \ref{Fig:Probe_pos} shows the position of the OLP compared to the plasma equilibrium for pulse 89344.
The proximity of the probe to the separatrix varies by less than $5$mm between pulses and can be considered fixed for this study. The probe is swept from $50$ to $-200$V on a $10$ms timescale, followed by a $20$ms plateau where the probe remains in ion-acceptance mode. Data during the voltage sweep is removed and only the central $15$ms of each $J_{sat}$ plateau is retained for analysis. This provides time-series' with 100,000 individual samples and approximately 1000 events over a threshold of $2.5$ times the standard deviation of the signal.
\\The original and usual role of the OLP is to measure steady state plasma fluxes to the wall, so they have not been designed with a view to making statistical measurements. This is reflected in the relatively modest $80$kHz sampling frequency of the probe. Filaments are often measured with characteristic timescales in the range of $10$s of $\mu$s \cite{D'IppolitoReview} which calls into question the appropriateness of the OLP for statistical measurements of filaments. To address this issue synthetic data-sets representative of measurements made by the OLP have been produced using the method of Kube \emph{et al}\cite{KubePoP2015}. The signal is composed of $N_{f}$ individual instances of a filament wave-form, taken here as the double exponential function 
\begin{equation}
f\left(t\right) = \begin{cases}
\exp\left(\frac{\left(t - t_{0}\right)}{0.1\tau_{d}}\right), & t < t_{0} \\ \\
\exp\left(-\frac{\left(t - t_{0}\right)}{\tau_{d}}\right), & t > t_{0} 
\end{cases}
\end{equation}
seeded randomly in time with a uniform distribution such that they represent a Poisson process. Their amplitudes are drawn from an exponential distribution. A Gaussian noise is added to the signal at a variable level. Here this was fixed at 10\%.
The synthetic signal adheres to the stochastic shot-noise model of Garcia \cite{GarciaPRL2012}. The underlying probability distribution function (PDF) of this model is a Gamma distribution characterized by the shaping parameter $\gamma = \tau_{d}/\tau_{w}$ where $\tau_{d}$ is the duration of the waveform and $\tau_{w} = N_{s}/N_{f}$ with $N_{s}$ the total number of samples. Two data sets have been produced, spanning $1$s in time and keeping $\gamma = 2.95$ fixed but varying $\tau_{d}$ and $\tau_{w}$ reciprocally.
These data sets are next sampled at a frequency of $1$MHz and $80$kHz respectively. Three key measurements are made and compared at each sampling frequency. These are: The PDF of the signal, the conditionally averaged wave-form of events above $2.5\sigma$ of the mean and the waiting time distribution of these same threshold events. The results of these measurements on the synthetic signals are shown in figure \ref{Fig:synth_pdfs} for filament wave-forms with duration times of $\tau_{d} = 2\mu$s, $20\mu$s and $200\mu$s.
\\Figure \ref{Fig:synth_pdfs} shows that the measurement of the PDF shape of the signal is unaffected by the sampling frequency used, even in the case where the filament waveform duration time, $\tau_{d}$, is well below the sampling time. The conditionally averaged profiles are also reproduced well in most cases. The exception is in the case where the 80kHz sampled signal cannot reproduce the $\tau_{d} = 2\mu s$ waveform. This is to be expected, however an encouraging trend is apparent with the measured waveform tending towards a point-wise delta function. This may be a useful tool in assessing whether the signal is under sampled or not. Finally the filament waiting time distribution is less well recovered in the 80kHz sampling case. For filaments with $\tau_{d} = 2\mu s$ the waiting time distribution is much wider than expected. This is also true for filaments with $\tau_{d} = 20\mu s$ however the degree of widening has reduced, and in the case of $\tau_{d} = 200\mu s$ the waiting time distribution is faithfully recovered. Despite the inability to reproduce the exact distribution, it should be noted that the trend in the distributions as $\tau_{d}$ increases is reproduced, ie a flattening of the PDF gradient as $\tau_{d}$ increases. As a result we may consider analysis of the trend of the waiting time distribution to be representative, but quantitative measurement of the distribution to be compromised by the relatively slow 80kHz sampling frequency for filament bursts with $\tau_{d}<20\mu$s.
\\Finally we note that as the flux of plasma to the probe increases we observe a minimal number of negative spikes in $J_{sat}$. These are likely to be the result of arcing of the OLP through the neutral gas that surrounds the probe as a result of recycling. As the plasma flux to the wall increases, the gas density increases and conductivity of the medium surrounding the probe is raised. The arc results in singular negative values in the signal which occupy, at maximum 3\% of the signal, but typically account for less than $0.5$\% of the signal. In the data presented herein we have removed these negative values, along with a window of two points either side of the negative value. All results have been computed with and without this removal and no significant changes to the results presented occur, and consequently the presence of these negative values has no impact on the conclusions we draw.
\section{Results}
\label{Sec:Res}
\begin{figure}[htbp]
\centering
\includegraphics[width=0.85\textwidth]{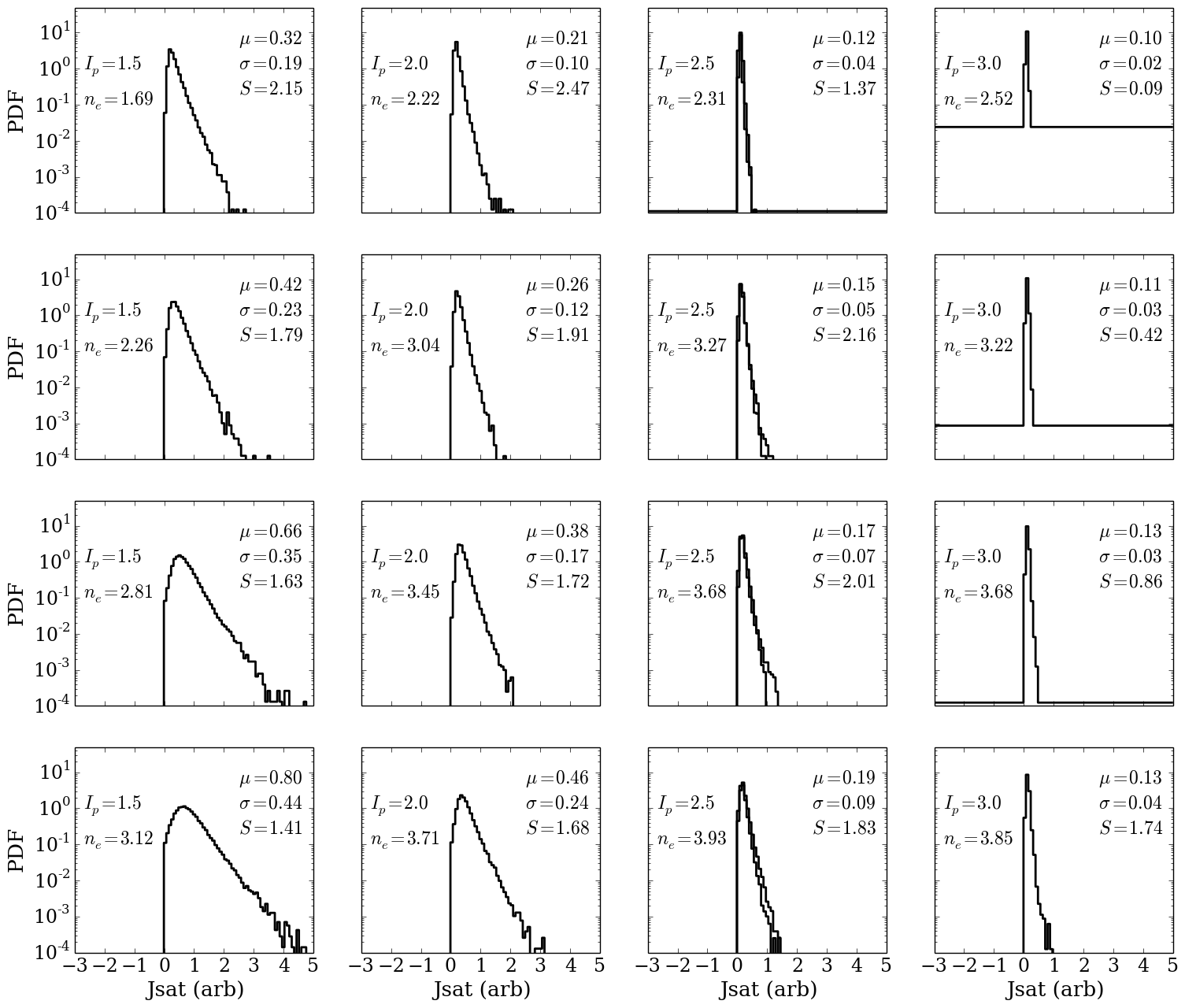}
\caption{PDFs of $J_{sat}$ measured by the OLP for each time window in each shot included in this study. Also given are the mean, $\mu$, standard deviation, $\sigma$, and skewness $S$ of the PDF for each case.}
\label{Fig:All_PDFs}
\end{figure}

\begin{figure}[htbp]
\centering
\includegraphics[width=0.7\textwidth]{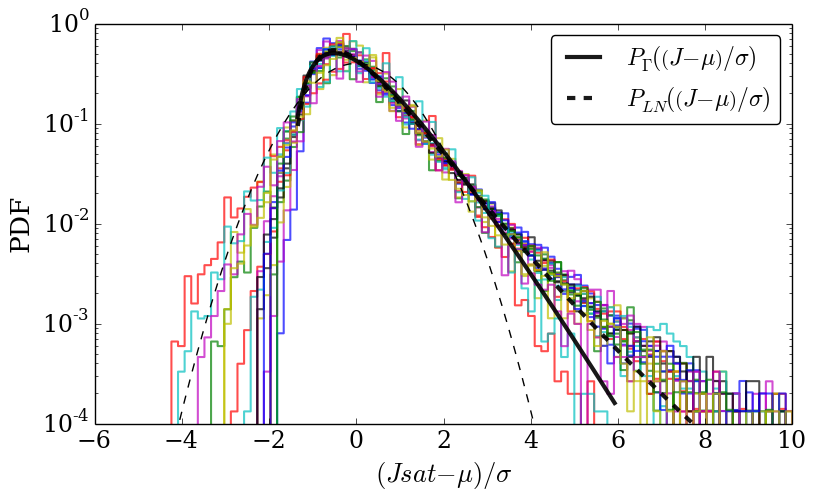}
\caption{Renormalized PDFs taken for each time window in each pulse studied. In total there are 20 individual experimental PDFs shown. A Gaussian distribution is overlaid for reference (thin dashed black line) alongside the best fit of a $\Gamma$ distribution (thick solid black line) and the best fit of a log-normal distribution (thick dashed black line).}
\label{Fig:All_PDFs_renorm}
\end{figure}

\begin{figure}[htbp]
\centering
\includegraphics[width=0.7\textwidth]{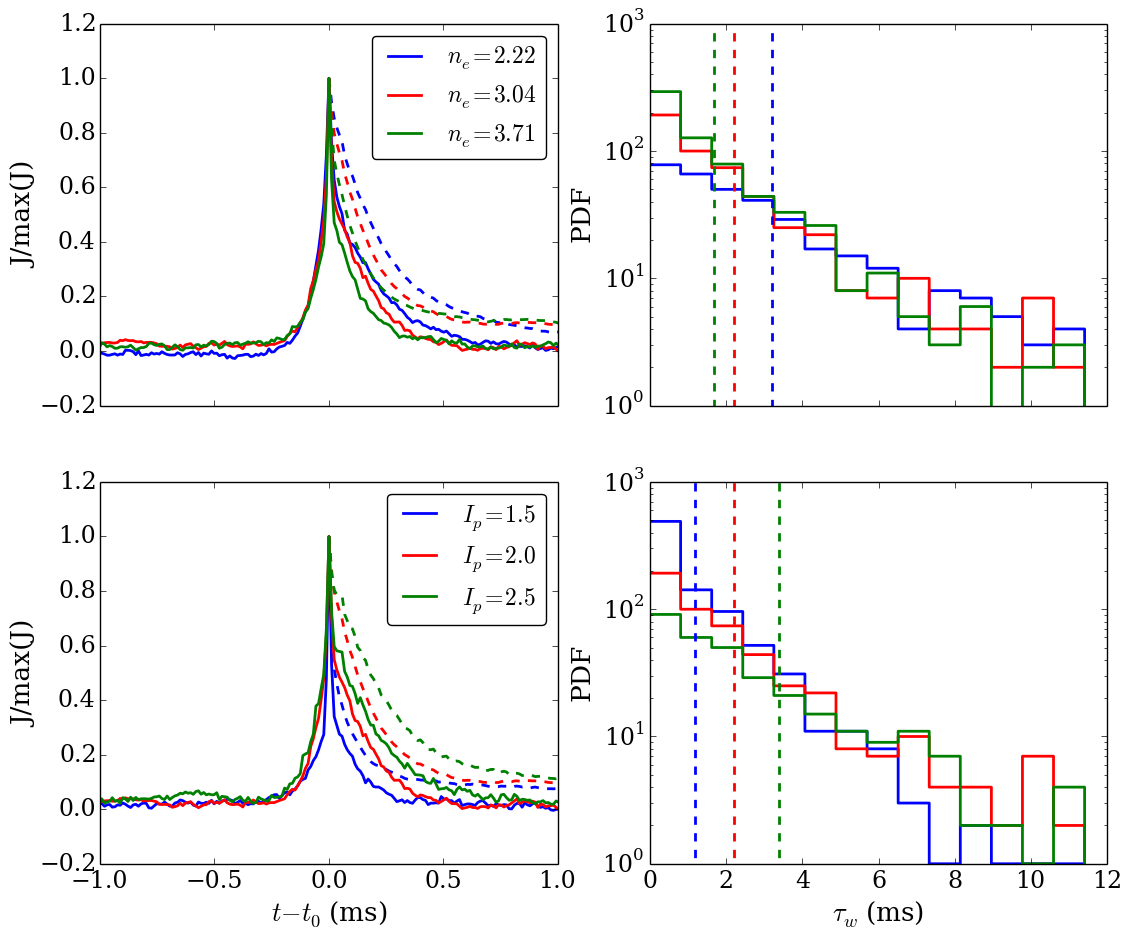}
\caption{Conditionally averaged $J_{sat}$ wave-form (left column) and distribution of burst waiting times (right column) for bursts with amplitudes satisfying $(J - \mu)/\sigma > 2.5$. Shown is a case where $I_{p} = 2$MA is held fixed (upper row) and with $n_{e}\approx 3.2\times 10^{19}m^{-3}$ held fixes (lower row). The autocorrelation time is shown (dashed lines, left column) and is comparable with the conditionally averaged waveform. The average waiting time (dashed line, right column) is shown to indicate clearly the trends in the waiting time istributions.}
\label{Fig:Cond_av}
\end{figure}

Figure \ref{Fig:All_PDFs} shows the PDFs for each of the time-series described in the previous section. 
\\The PDFs of the raw signal show significant variation with the following trends evident: The mean, $\mu$, and standard deviation, $\sigma$, of the signal increase with line averaged density and decrease strongly with plasma current. The ratio $\sigma/\mu$ varies in value from $0.2$ to $0.59$ across the data set with the lowest values occuring in the $I_{p}=3.0$ time-series. As noted by Kube \emph{et al}\cite{KubePoP2015} measurement of higher order moments of the PDF are subject to significant variation due to inherent statistical variation. In addition the presence of a background random noise in the signals also provides a source of variation. By comparing against synthetic signals we have attributed the variation in $\sigma/\mu$ and skewness, $S$, to inherent statistical uncertainty and random noise, whilst the variation in $\mu$ and $\sigma$ cannot be accounted for by these processes and are therefore considered robust effects.  We note that the increases in $\mu$ and $\sigma$ occur concomitantly with a broadening of the SOL density profile, however we reserve any analysis of this occurance for the future, focussing solely on single-point measurements made at the wall. 
\\The shape of the PDF can be compared upon renormalization achieved by subtraction of $\mu$ and rescaling by $\sigma$. After renormalization we find that all PDFs collapse approximately onto a single curve. This collapse is illustrated in figure \ref{Fig:All_PDFs_renorm}. 
We note that the PDFs do not perfectly collapse. This is in accordance with the variation observed in $\sigma/\mu$ and $S$ over the data set. Indeed a perfect collapse would also imply the constancy of these two parameters. As noted before, this deviation is the result of statistical uncertainty and random noise which affects, in particular, the tails of the distribution where the number of samples decreases. In signals with a low signal to noise ratio, as in the data from pulse $89351$ with $I_{p} = 3$MA, random noise causes the PDF to tend more towards a Gaussian as can be seen in figure \ref{Fig:All_PDFs_renorm}. As such these time-series can be considered outliers and are rejected from subsequent analysis. The variation in all other PDFs can be attributed to statistical and random noise and has been validation using synthetic data. For this reason the rest of the data presented in figure \ref{Fig:All_PDFs_renorm} can be considered to have collapsed.
\\We have compared the experimental distribution to both a $\Gamma$ distribution, described by
\begin{equation}
	P_{\Gamma}(x,\gamma) = \frac{x^{\gamma - 1}}{\Gamma\left(\gamma\right)}\exp\left(-x\right)
\end{equation} 
and a log-normal distribution given by
\begin{equation}
	P_{LN}(x,\lambda) = \left(\lambda x \sqrt{\pi} \exp\left(-\frac{1}{2}\left(\frac{\ln(x)}{\lambda}\right)^{2}\right)\right)^{-1}
\end{equation}
In the positive tail of the distribution the PDF exhibits a higher probability of occurrence than predicted by the $\Gamma$ distribution and the log-normal distribution provides a better fit with a RMS relative error of 0.32 compared to 0.92 for the $\Gamma$ distribution. We have not included the aforementioned cases at $I_{p} = 3$MA where the signal-to-noise ratio is low in the fit. For the $\Gamma$ distribution we find a shaping parameter $\gamma = 2.956$ whilst for the log-normal distribution we find a shaping parameter of $\lambda = 0.458$. The deviation in the tail of the PDF is a common observation in the limiter shadow of devices and has been previously observed qualitatively on both Langmuir probes \cite{GarciaJNM2007,LaBombardPoP2001,LaBombardNF2005,BoedoPoP2003} and gas-puff imaging \cite{GarciaJNM2013,GarciaPoP2013}. 
\\The PDFs collapse over approximately four decades of the ordinate. This feature, alongside the large variation in plasma parameters over which this study was conducted is strong evidence of universality in the ion flux to the JET wall. Garcia's stochastic model \cite{GarciaPRL2012} links the shaping parameter of the PDF to the intermittency parameter such that $\gamma = \tau_{w}/\tau_{d}$ where $\tau_{d}$ is the duration time and $\tau_{w}$ is the waiting time of the bursts that form the signal. The collapse of the PDF observed here implies that $\gamma$ is a constant across significant variation in line-averaged density and plasma current. We note here, and elaborate further in the discussion, that this is a potentially important observation for the prediction of first-wall fluxes. $\gamma$ can be held constant through two possible mechanisms: Both $\tau_{d}$ and $\tau_{w}$ are constant over the parameter space or they balance such that their ratio is constant. These two possibilities have been disambiguated by identification of burst events with an amplitude $(J - \mu)/\sigma > 2.5$. In figure \ref{Fig:Cond_av} we have presented the conditional averages and waiting time distributions of these thresholded events for a scan in $n_{e}$ at constant $I_{p} = 2.0$MA and for a scan in $I_{p}$ with $n_{e}$ in the range $[3.04,3.27]\times 10^{19}m^{-3}$. Also shown is the autocorrelation time and the average waiting time between events. 
\\As either $I_{p}$ increases or $n_{e}$ decreases there is a widening of the conditionally averaged wave-form (and correspondingly an increase in the autocorrelation time) which occurs alongside an increase in the average burst waiting time. This indicates that the invariance of the PDF observed as $I_{p}$ and $n_{e}$ vary is the result of a balance between the temporal width of the burst on the probe and the frequency of bursts hitting the probe.  We note that these results are consistent with both a plasma current scan \cite{GarciaPPCF2007} and density scan \cite{GarciaNF2007} in TCV. We also note though that Carralero \emph{et al} report \cite{CarraleroNF2014} an increase in the autocorrelation time of fluctuations in the ASDEX-Upgrade SOL as density is increased. This increase is marginal in the far-SOL, and with the study being carried out with $600$kW of external heating the scenario is not directly comparable with the Ohmic plasmas here. Furthermore the data used here contains an order of magnitude more bursts udring the sampling time, providing robust data for statistical analysis. Extending this OLP analysis to externally heated plasma in JET may help to explain this disparity.

\section{Discussion and Conclusions}
\label{Sec:Conc}
We have presented analysis of wall-mounted langmuir probe signals of ion saturation current in JET over a significant range of plasma densities and plasma currents in Ohmically heated L-mode plasmas in the horizontal target configuration. We find that the mean and standard deviation of the probe signal increases with line-averaged density or with a decreasing plasma current. Despite this variation, after subtraction of the mean and rescaling by the standard deviation of the signal, the PDFs of all the time-series collapse onto a single curve. The shape of this curve is better fit by a log-normal distribution than a $\Gamma$ distribution, which is commonly found to be the case in the limiter shadow of tokamak devices. Despite the collapse of the PDFs, we have shown by taking conditional averages of bursts within each time-series that the underlying temporal structure of the bursts vary as plasma current and density vary. In particular an expansion in the temporal width of the burst wave-form is observed with increasing plasma current and decreasing density. In order for the PDFs to remain invariant this requires that the frequency of bursts must adapt to compensate the change to the burst wave-form and this is verified by measurement of the burst waiting times. Since the measurements taken are single point measurements of events with both a spatial scale and a velocity normal to the probe it is not possible to assess for certain whether the change to the burst wave-form is the result of a change in the spatial structure of the bursts, or a change in their velocity towards the probe. To properly decouple the effect of shape and velocity a similar study should be conducted using either multiple but radially separated probes, or using a 2D measurement such as Li-BES \cite{BrixRSI2012}. Even without decoupling this effect we note that prediction of the observed invariance of the PDF and the balance between burst duration time and waiting time observed here may be a good metric for determining the suitability of non-linear models at capturing the dynamics of SOL turbulene. Furthermore the constraint that the intermittency parameter remains constant may be an important factor to consider when predicting first wall ion fluxes for future machines. 
\\Finally we note that this study is limited in plasma configuration to Ohmically heated horizontal target plasmas. To further assess the universality of the plasma wall flux in JET this study should be extended to different plasma shapes, divertor configurations and heating mechanisms as well as an extension into H-mode.

\section{Acknowledgements}
This work has been carried out within the framework of the EUROfusion Consortium and has received funding from the Euratom research and training programme 2014-2018 under grant agreement No 633053. The views and opinions expressed herein do not necessarily reflect those of the European Commission.

\section{References}
\bibliographystyle{prsty}
\bibliography{Bibliography}

\end{document}